\documentstyle[twoside,fleqn,espcrc2,psfig]{article}

\def\lsim{\raise0.3ex\hbox{$\;<$\kern-0.75em\raise-1.1ex\hbox{$\sim\;$}}}
\def\gsim{\raise0.3ex\hbox{$\;>$\kern-0.75em\raise-1.1ex\hbox{$\sim\;$}}}

\newcommand{\bea}{\begin{eqnarray}}
\newcommand{\eea}{\end{eqnarray}}

\def\beq{\begin{equation}}
\def\eeq{\end{equation}}

\def\bea{\begin{eqnarray}}
\def\ba{\begin{array}}
\def\ea{\end{array}}

\def\lsim{\;
\raise0.3ex\hbox{$<$\kern-0.75em\raise-1.1ex\hbox{$\sim$}}\;
}

\newcommand{\AmS}{{\protect\the\textfont2
  A\kern-.1667em\lower.5ex\hbox{M}\kern-.125emS}}

\title{
\begin{flushright}
\vglue -3.cm
   {\small
    DF/IST-5.98  \hglue 2.cm  \\
    FISIST/13-98/CFIF/CENTRA\hglue 2.cm \\
   }
\end{flushright}
Neutrino Magnetic Moment and Solar Neutrino Experiments 
}

\author{Ana M. Mour\~ao
        \address{CENTRA - Centro Multidisciplinar de 
                 Astrof\'{\i}sica
                 and Dep. of Physics, I.S.T.\\
                 Avenida Rovisco Pais, 1, 1096 Lisboa Codex - Portugal}
        \thanks{This work was in part supported by JNICT  projects
         PRAXIS/PCEX/P/FIS/4/96 and  ESO/P/PRO/1127/96, Funda\c{c}\~ao Oriente
         and GTAE-Lisbon. }
and
         Anna Rossi\address{CFIF - Centro de F\'{\i}sica das Interac\c{c}\~oes 
         Fundamentais  - I.S.T. \\
        Avenida Rovisco Pais, 1, 1096 Lisboa Codex - Portugal
        } 
\thanks{Supported by a grant from GTAE - Lisbon. 
}}
\begin{document}
\renewcommand{\thefootnote}{\fnsymbol{footnote}}

\begin{abstract}

 We have studied the effect of a non-vanishing neutrino 
 magnetic moment ($\mu_{\nu}$) on the $\nu_{\rm x}$
  ({x=$e,\mu,\tau$}) elastic scattering off electrons for
 the Super-Kamiokande detector. 
The bounds on the $\mu_{\nu}$ we have obtained are comparable to 
that extracted from laboratory experiments.
Furthemore, we outline the potential  of the Borexino 
experiment which may be sensitive to neutrino magnetic moments  
$\lsim 10^{-10}\mu_B$.  In our analysis we have considered 
both cases of Majorana and Dirac neutrinos.

\end{abstract}
\maketitle
\renewcommand{\thefootnote}{\arabic{footnote}}
\setcounter{footnote}{0}
\noindent
{\bf 1. Introduction}

\vspace{0.3cm}

  The solar neutrino problem (SNP) is nowadays regarded as 
   a direct evidence for  physics beyond the standard electroweak model.
  This is due to the fact that the observed
  deficit of electron  neutrinos  in all solar neutrino experiments
  can only be explained assuming that  non-zero neutrino masses 
  and/or neutrino magnetic moments might lead to flavour, spin
  or spin-flavour neutrino oscillations of the  solar left-handed 
  neutrinos 
  $\nu_{e\rm L}$ \cite{CDF}-\cite{MSW}.  At the same time 
  a lot of work has being done also to understand the 
  implications of the uncertainties in  helioseismology \cite{seismo}
  and nuclear physics \cite{nphys}  for the SNP. 
  
   As was already pointed out in the literature \cite{suz91}-\cite{PM98}
    a non-vanishing  neutrino magnetic moment, $\mu_{\nu}$, 
   can also affect the neutrino elastic scattering off electrons  
   through which solar $\nu$'s 
   are detected in Super-Kamiokande (SK).  
   Therefore the expected signal in  such a detector may depend 
   also on the electromagnetic  properties of the neutrinos.
     
     In previous works  \cite{MPR92,PM98} bounds have been obtained on 
   $\mu_\nu$
    from Kamiokande and Super-Kamiokande data, taking into 
    account the restrictions on
    the  $\nu_e$ survival probability imposed by  Homestake,  
    Gallex and  SAGE experiments.

     In this work we present an updated analysis
     of the effect of a non-vanishing  magnetic
     moment  on the scattering off electrons for the Super-Kamiokande
     detector, similarly to the study carried on  in\cite{PM98}.
     We have considered both the cases of  Dirac and 
     Majorana  neutrinos.

      For definiteness  we examine the two neutrino system
      $\nu_e - \nu_{\rm x}$  (x=$\mu , \tau$) with a non-zero mass 
       difference $\delta m^2$.
       In the solar interior the spin-flavour resonant conversion 
       \cite{RSFP}
       $\nu_{e \rm L}\rightarrow \nu_{\rm x R}$,  with probability
       P$_{\rm R}$,  occurs at higher matter density with respect to
       the usual MSW resonant conversion 
       $\nu_{e \rm L}\rightarrow \nu_{\rm x L}$ \cite{MSW},
characterised by the probability P$_{\rm L}$\footnote{We remind that 
the $\nu_{\rm x L}\rightarrow \nu_{e\rm  R}$ spin-flavour resonance 
would occur at a much higher matter density, for given $\delta m^2$. 
We assume therefore that it does not take place  in the solar interior.}.
As a result of both conversions we have three different neutrino
`flavours' reaching the Earth:
\bea 
\Phi_{\nu_{\rm x R}}\!\!\!\!&=\!\!\!\!&{\rm P}_{\rm R} \,\Phi_{\rm SSM}~,
\nonumber\\
\Phi_{\nu_{\rm x L}}\!\!\!\!&=\!\!\!\!&{\rm P}_{\rm L}(1- {\rm P}_{\rm R})
\, \Phi_{\rm SSM}\equiv {\rm P}_{\rm x L}\Phi_{\rm SSM}~,\label{fluxes}\\
\Phi_{\nu_{e\rm  L}}\!\!\!\!&=\!\!\!\!&(1- {\rm P}_{\rm L})
(1-{\rm P}_{\rm R}) \,\Phi_{\rm SSM}
\equiv {\rm P}_{e\rm  L}\,\Phi_{\rm SSM}~\nonumber,
\eea
\noindent
here $ \Phi_{\rm SSM}$ is the standard solar  model (SSM) prediction 
for a certain component of the neutrino flux \cite{BP98}. 
In the above relations
\footnote{The picture envisaged from the eq. (\ref{fluxes}) 
could appear, for example, when  
the MSW and spin-flavour resonances lie far away 
from each other to be treated separately. This may be the case 
in the central region of the sun.} 
there are only two independent quantities, 
{\em e.g.}   ${\rm P}_{e\rm  L}$  and   ${\rm P}_{\rm  R}$.
Hence  ${\rm P}_{\rm  L}= 1-\frac{{\rm P}_{e\rm  L}}{1-{\rm P}_{\rm  R}}$
where   ${\rm P}_{\rm  R}\leq 1- {\rm P}_{e\rm  L}$.
On the basis of this picture we have calculated the expected signal
 in SK experiment.

For the sake of simplicity, we assume that                       
all  the  probabilities are energy independent.
            In particular the present  SK data on the
            $\nu_{e\rm L}$ energy spectrum do 
not yet exclude that 
            ${\rm P}_{e\rm  L}$  is energy independent for
            E$_\nu \geq 6.5$ MeV \cite{sk98}.

\vspace{0.5cm}
\noindent     
{\bf 2. Signal in the Super-Kamiokande detector}
 
\vspace{0.3cm}
     
   The total signal in the Super-Kamiokande experiment, for the case of
   Dirac neutrinos,  can be written as 
\vglue -0.5cm   
\bea
R_{SK}^{\rm total}&=&R_{SK}^{\rm w + em} \nonumber \\
            ~&=&P_{e \rm L} \left[ \langle \sigma ^{\rm w}_{\nu _{e \rm L}}\rangle  +   
       \langle \sigma ^{\rm em}_{\nu _{e \rm L}}\rangle    \right] \nonumber \\
 ~     &~&+P_{\rm xL} \left[ \langle \sigma ^{\rm w}_{\nu _{ \rm x L}}\rangle +
      \langle \sigma ^{\rm em}_{\nu _{ \rm x L}}\rangle    \right] \nonumber \\
&~&+P_{\rm R}  \langle \sigma ^{\rm em}_{\nu _{ \rm x R}}\rangle 
\eea
 
\noindent
where the averaged total $\nu- e$ cross sections are 
\beq
<\sigma^{i}_{\alpha}>=\int dE_{\nu}
\Phi^{^8 B}_{\rm SSM}(E_\nu)\sigma ^{ i}_{\alpha}(E_{\nu}), 
\eeq 
\noindent
where $\alpha= \nu_{e L},\nu_{ \rm x L,R}$  and 
$i$=em for electromagnetic  and $i$=w for weak cross sections. 
~$\Phi^{^8 B}_{\rm SSM} (E_\nu)$ is the 
$^8B$ solar neutrino flux from BP98 \cite{BP98}.
In the calculation of  the cross sections $\sigma^{ i}_\alpha (E_{\nu})$ 
we have taken into account the energy resolution of the detector 
\cite{JB}.
The electromagnetic cross section can be taken e.g. from
Kerimov {\it et al} \cite{KSN}.  
Taking into account that 
$ <\!\sigma^{\rm em}_{\alpha}\!>\,\propto \mu_{\nu}^2$
it is easy to see that  
for  $P_{e \rm L}=1, P_{\rm R}=P_{\rm x L}=0$, and $\mu_{\nu _e}=0$ 
we obtain the SSM expectation for the signal. 
 
In our analysis we use the most    
recent   SK data for the $^8$B solar neutrino 
flux,  $\Phi_{\rm SK}^{exp}=(2.44\pm 0.05)\times 10^6
{~\rm cm}^2{\rm s}^{-1}$ \cite{suzukinu98}, 
normalised to the SSM prediction - 
$\Phi_{\rm BP98}^{th}=(5.15\pm 0.98) \times 10^6{~\rm cm}^2{\rm s}^{-1}$
\cite{BP98},  namely  
 \beq
 Z_{\rm K}=\frac{\Phi_{\rm SK}^{exp}}{\Phi_{\rm BP98}^{th}}=0.47\pm 0.09.
 \eeq

 As opposite to the Super-Kamiokande experiment, the Homestake detector is
 only sensitive to the $\nu_{e\rm L}$ component of the solar neutrino 
 flux and the total rate is mainly due to $^8$B neutrinos. 
 Considering the experimental data \cite{clorine} $Z_{\rm Cl}=0.28\pm 0.03$,
 we can assume that for the higher energy spectrum of $^8$B neutrinos, 
  E$_\nu \geq 6$ MeV, the neutrino survival probability 
is $P_{e\rm L}\sim 30$\%. Therefore this implies a 
  total depletion of  the intermediate energy  $^7$Be  neutrinos
  as the present understanding of the SNP points to
 \cite{CDF}-\cite{BKS98}.
 
 We can now find the values of the neutrino magnetic moment
 compatible with solar neutrino experiments, i.e needed
 to obtain the signal observed at the Superkamiokande.
 
 \vspace {0.5cm}
 \noindent
 {\bf 3. Limits on neutrino magnetic moments} 

 \vspace {0.3cm}

 We have studied the impact of non-vanishing $\mu_{\nu}$ in the SK signal
 taking into account the contribution of the several neutrino `flavours'
 as shown in (\ref{fluxes}). In Fig. 1 we show the contour-plots for 
 $Z_{\rm K}$ =0.47 in the parameter space $(P_{\rm R}, \mu_\nu)$, where
 $\mu_\nu\equiv\mu_{\nu_e}=\mu_{\nu_{\rm x}}$. We repeat our analysis
 with other  values of  $Z_{\rm K}$, 
 just to understand implications of  uncertainties in the SSM used
 in the evaluation of  $\mu_{\nu}$.
  Notice that the larger the value of 
   $P_{\rm R}$ the larger the value of $\mu_\nu$ needed to 
   saturate $Z_{\rm K}$ in order to compensate for the loss 
   of the $\nu_{\rm x L}$ component (recall that  P$_{e \rm L}$ 
is fixed at 0.3)\,.
   The present experimental $Z_{\rm K}$  implies the bound 
$\mu_{\nu} \lsim (2\div 5) \times 10^{-10} \mu_{\rm B }$, 
 almost independently of    $P_{\rm R}$.

We have considered also the  case with $\mu_\nu\equiv\mu_{\nu_{\rm x}}$ and 
$\mu_{\nu_e}=0$.  In this case we obtain a similar plot as that in Fig.1 
and the limit 
 $\mu_{\nu} \lsim 5 \times 10^{-10} \mu_{\rm B }$.

  For the sake of completeness we have also studied the case of 
  Majorana neutrinos, for which the {\it antineutrino} state  
  $\nu_{\rm x R}= \nu^c_{\rm x L}\equiv\tilde{\nu}_{\rm x}$, is `active', 
having  both electromagnetic and standard weak interactions.  
Hence in the eq. (2)
  one more term is to be added, {\it i.e.}   
  $P_{\rm R}\langle\sigma_{\tilde{\nu}_{\rm x}}^{\rm w}\rangle$.
   We remember that for  Dirac neutrinos
    both diagonal or transition $\mu_{\nu}$
     can be generated \cite{FS}, while  for Majorana neutrinos 
   only transition (off-diagonal) magnetic moments are allowed
   \cite{SV81}.   

In Fig. 2 we show our result in this scenario: 
the upper bound  for Majorana neutrinos comes out to be similar,
$\mu_{\nu} \lsim (1.5\div 4.0) \times 10^{-10} \mu_{\rm B }$
 for the extreme case $P_{\rm R}=0.7$ ~. Note that 
for   $P_{\rm  R} < 0.7~$, smaller values of $\mu_{\nu}$ are in principle 
tolerated as the $\tilde{\nu}_ {\rm x}$'s contribute to the weak cross 
section (compare with the Dirac case shown in Fig. 1).
Our upper limit was obtained assuming vanishing vacuum 
mixing angle (and then P$_{\rm L}=0)$, thereby satisfying 
the experimental constraints on the $\!\tilde{\nu}_{e}\!\!\!\!$
 which could emerge
from the subsequent vacuum oscillation 
$\tilde{\nu}_{\rm x}\rightarrow\tilde{\nu}_{e}$\cite{antinue}.

We conclude that  the bounds we have obtained

\newpage
\vglue -15mm
\begin{figure}[ht]
\hglue -9mm
\psfig{file=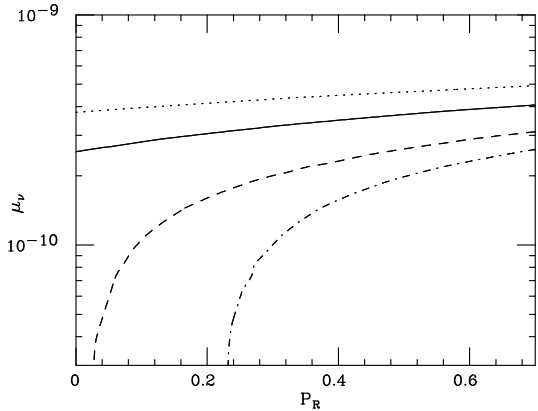,width=90mm,angle=90}
\vglue -11mm
\caption{The contour plots of the expected signal versus the SSM prediction 
$Z_{\rm K}$ in Super-Kamiokande in 
the (P$_{\rm R}~, \mu_\nu$) parameter space in the case of Dirac neutrinos. 
The dotted, solid,  dashed and dot-dashed curves correspond to 
$Z_{\rm K} = 0.55, 0.47, 0.4$ and $Z_{\rm K}= 0.37$, respectively.
$\mu_\nu$ is given in units of Bohr magneton.}
\vglue -5mm
\end{figure}

\noindent 

are slightly more restrictive than  those from  
 accelerator experiments, namely in the case of $\mu_{\nu_\mu}$ it is 
$~\mu_{\nu_\mu} < 7.4 \times 10^{-10}  \mu_B$\cite{munuexp}. 
However our bounds are still not comparable to those from reactor  experiments, -    
$\mu_{\nu_e} < 1.8\times 10^{-10} \mu_B$.  Needless to say that our results 
are more stringent in the case of $\mu_{\nu_\tau}$ for which 
 $\mu_{\nu_\tau} < 5.4\times 10^{-7} \mu_B$ \cite{RPP}.


\vspace{0.3cm}

Finally, we have discussed the potential of the future Borexino experiment
\cite{borex} which will detect   $^7$Be neutrinos through $\nu-e$ 
elastic scattering. 
For $\mu_\nu = 10^{-10} \mu_B$, the cross section 
$\sigma^{em}$ can be comparable to 
 $\sigma^{\rm w}_{\nu_e}$ for $E_\nu\leq 1$ MeV. Therefore
 we can expect a substantial 
signal in Borexino even in the case of complete conversion  of the initial
$^7$Be-$\nu_e$'s into  ${\nu}_{\rm x R}$ or $\tilde{\nu}_{\rm x}$.
In Fig. 3 we have plotted 
the energy distribution of the events for Borexino 
in the case of complete conversion $\nu_e\rightarrow \tilde{\nu}_{\rm x}$ 
(dotted line). We note that in  this distribution we have 
taken into account 
the contributions 
from all solar  neutrino
components. However the 
$^7$Be 
neutrinos contribute to  more than 90\% of the signal. 
Other important contribution is given by the {\it pep} flux
\newpage
\vglue -1.5cm
\begin{figure}[ht]
\hglue -9mm
\psfig{file=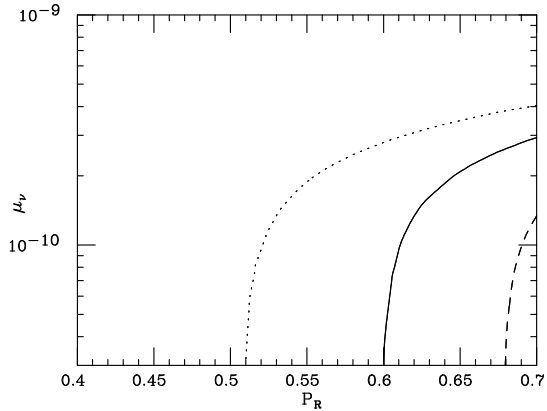,width=90mm,angle=90}
\vglue -1.1cm
\caption{The same as in Fig.1 but for the Majorana case.}
\vglue -0.5cm
\end{figure}

\noindent
 of solar neutrinos. 
For comparison we also shown the SSM distribution (solid line).
We can expect $\sim 50\%$ of the SSM prediction and a specific distortion 
of the spectrum. 
This is in contrast with the case of pure MSW conversion ($\mu_\nu = 0$) 
${\nu}_e\rightarrow {\nu}_{\rm x}$ that would imply a $(20\div 25) \%$ 
reduction in the signal.

\vspace*{0.5cm}
\noindent
 {\bf 4. Conclusions}

\vspace{0.3cm}

In this contribution we have updated the analysis on the effect of a 
non-vanishing neutrino magnetic moment on the $\nu-e$ cross section 
in Super-Kamiokande experiment.
The limits we achieved -$\mu_\nu\lsim (2\div 5) \times 10^{-10}\mu_B$ -
remain comparable to that extracted from the previous Kamiokande data  
as the electromagnetic cross section is smaller than the weak one
 for the energy range involved, $E \geq 6$ MeV.
Therefore experiments with a much lower energy threshold - such as 
  Borexino 
or Hellaz - could exhibit a
much better sensitivity to a non-zero 
$\mu_\nu$ and consequently
provide a better testing of the 
spin-flavour reso\-nant   conversion itself as a solution to 
the SNP\cite{borex,Pastor98}.
 \newpage
 
\vglue -1.5cm
\begin{figure}[h]
\hglue -9mm
\psfig{file=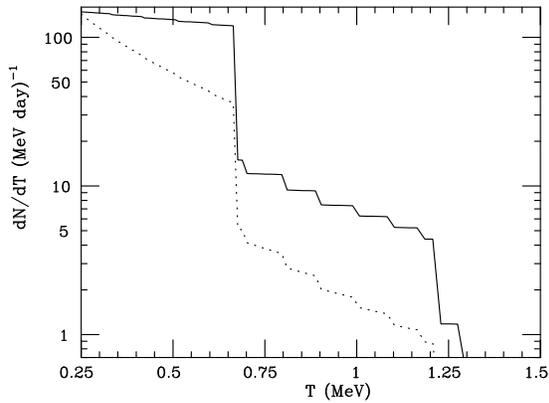,width=90mm,angle=90}
\vglue -1.1cm
\caption{The energy distributon of the events in Borexino experiment. 
The dotted curve represents the expected signal in the case of complete 
$\nu_e\rightarrow \tilde{\nu}_{\rm x}$ (Majorana case) with 
$\mu_\nu = 10^{-10} \mu_B$. The solid line corresponds to the SSM 
distribution.}
\vglue -0.5cm
\end{figure}

 {\bf Acknowledgments}

\vspace{0.3cm}
 
We are grateful to  E. Akhmedov, L. Bento, D. Gough and  Y. Suzuki 
for very useful  discussions. 
One of us (A.M.) would like to thank
the organisers  for 
the invitation to present this work and for 
the very stimulating  $\nu 98$ Conference.

\end{document}